\documentstyle[12pt,moriond,psfig]{article}
\def\spose#1{\hbox to 0pt{#1\hss}}
\def\lta{\mathrel{\spose{\lower 3pt\hbox{$\mathchar"218$}}
     \raise 2.0pt\hbox{$\mathchar"13C$}}}
\def\gta{\mathrel{\spose{\lower 3pt\hbox{$\mathchar"218$}}
     \raise 2.0pt\hbox{$\mathchar"13E$}}}
\def\etal{{\it et al.\ }}

\newcommand{\R}{${\cal R}$}

\begin{document}
\heading{CHEMICALLY  CONSISTENT  EVOLUTION  OF  GALAXIES  ON COSMOLOGICAL  TIMESCALES  AND  THE DLA  GALAXY  POPULATION}

\author{U. Fritze -- v. Alvensleben, U. Lindner, C. S. M\"oller}
{Universit\"atssternwarte G\"ottingen, Germany.}
{\ }

\begin{moriondabstract}
We describe the evolution of galaxies in a 
chemically consistent way accounting for the increasing initial 
metallicity of successive generations of stars. The enrichment of various types of
model galaxies is compared with Damped Ly$\alpha$ ($=$ DLA) abundances 
over the redshift range from ${\rm z \sim 0}$ through ${\rm z > 4.4}$.
We discuss properties and composition of the galaxy population
giving rise to DLA absorption. Beyond evolution of individual
galaxies a change is found in the composition of DLA galaxy samples
from high to low redshift. Spectrophotometric properties are predicted for 
optical identifications of DLA galaxies. 
\end{moriondabstract}

\section{Introduction, Models and Input Physics}
Over the last few years evidence for broad metallicity distributions in galaxies has been 
accumulating as well as for the importance of subsolar abundances for late-type spirals, dwarf and LSB galaxies, in the local Universe and even more so during early evolution. 
Our chemically consistent evolutionary synthesis approach attempts to account for the increasing metallicity of successive stellar generations. We keep track of the enrichment history of the ISM and follow successive generations of stars by using stellar evolutionary tracks, element yields, lifetimes, spectra, etc. for a range of metallicities Z from ${\rm 10^{-4} ~ to ~ 0.05}$. Galaxies of various spectral types are 
described by appropriate star formation histories (SFHs). For Sa, Sb, and Sc spirals we use SFRs that are linear functions of the evolving gas-to-total-mass ratios with characteristic timescales for SF ${\rm t_{\ast}}$ -- defined via ${\rm \int_0^{t_{\ast}} \Psi~dt~=~(1 - \frac{1}{e}) G}$ -- increasing from 2 Gyr (Sa) to 10 Gyr (Sc). For Sd spirals we use a const. SFR. Being very simplified 1 -- zone descriptions without any dynamics included, our models are meant to describe the evolution of global quantities like integrated luminosities and colours, and average ISM abundances. 

Our first set of chemically consistent (hereafter {\bf cc}) models were presented for the photometric evolution by 
Einsel \etal 1995. The cc models were then extended to cover the spectral evolution of nearby galaxies in M\"oller \etal 1997. In Lindner \etal 1999 we present the chemical evolution aspects of our unified cc chemical and spectrophotometric evolutionary models and compare -- for a standard cosmology -- the redshift evolution of model ISM abundances with data on DLAs. In M\"oller \etal ({\sl in prep.}) we will present cc spectro-cosmological evolution models and compare to high-redshift galaxy observations. Here, we discuss our chemo-cosmological evolution models in comparison with DLA 
abundances over the redshift range ${\rm z \sim 0.4}$ to ${\rm z \gta 4}$, draw conclusions as to the nature of the DLA galaxy population and its redshift evolution and use the spectro-cosmological aspects of our cc models to predict optical and NIR properties of DLA galaxies. 

We use 5 different databases of input physics for 5 metallicities ${\rm Z=10^{-4}~ through ~Z=0.05}$. Model atmosphere spectra are from Lejeune \etal 1997, 1998, stellar evolutionary tracks and lifetimes are from the Padova group, remnant masses and stellar yields for elements H, He, C, N, O, Ne, Mg, Al, Si, S, Cr, Mn, Fe, Ni, Zn, ... from van den Hoek \& Groenewegen (1997) for stars with ${\rm m < 8 M_{\odot}}$ and from Woosley \& Weaver (1995) for stars ${\rm > 8 M_{\odot}}$. SNIa contributions from carbon deflagration white dwarf binaries are 
included as prescribed by Matteucci (see Lindner \etal 1999 for details). We start with the lowest metallicity, form stars continuously according to the SF laws above and follow the evolution of the stars in the HRD as well as their gas and heavy element production with stellar tracks and yields for the lowest metallicity. Once the ISM abundance reaches a certain threshold stars formed thereafter are described with a set of input physics for the next higher metallicity, and so on. 

Basic parameters of this kind of evolutionary synthesis models are the SFH and the IMF. 
For any cosmological model characterised by the parameters (${\rm H_o,~\Omega_o,~\Lambda_o}$) and an assumed redshift of galaxy formation ${\rm z_{form}}$, the ISM abundance evolution as a function of time directly transforms into a redshift evolution. For the spectro-cosmological evolution of apparant magnitudes and colours cc 
cosmological and evolutionary corrections and attenuation by intergalactic hydrogen are taken into account. 
The SFHs are chosen as to provide agreement -- for a Scalo IMF -- of the models after a Hubbble time with observed integrated average colours of nearby galaxies (RC3), template spectra (Kennicutt 1992), and characteristic HII region abundances as observed at ${\rm r \sim 1 R_e}$ (Oey \& Kennicutt 1993, Zaritsky \etal 1994, Ferguson \etal 1998) for the respective galaxy types. While ${\rm 1~R_e}$ seems a reasonable radius at which to compare HII region abundances with our global galaxy models, Phillipps \& Edmunds 1996 argue that this is also the most probable radius for an arbitrary QSO line of sight to cut through an intervening disk when producing damped Ly$\alpha$ absorption. 

\section{DLA Abundances and Absorber Properties: \\
Results and Discussion}
Abundances of elements in DLAs habe been measured from weak, non-saturated low ionisation lines (which do not require ionisation corrections) for quite a while. However, only since the velocity structure of these lines can be fully resolved in HIRES spectra precise abundances can be determined. As summarised by Pettini ({\sl this volume}) these high-quality data are now becoming available for many elements in a large number of DLAs all over the redshift range from ${\rm z \gta 4}$ down to ${\rm z \sim 0}$. Since DLA systems are not biased to high luminosity, SFR, or radio power -- they just happen to lie in front of a more distant QSO -- chemical information about ``normal gas-rich galaxies'' is available now over a look-back time of $> 90 \%$ of the Hubble time, much larger than the range accessible to spectrophotometric observations. 
Only a weak trend is seen in the redshift evolution of DLA abundances with a large intrinsic scatter in particular towaards higher redshifts. 

From the high column densities ${\rm log N(HI) \geq 20.5}$, a flat geometry (from statistical analyses) and the extent perpendicular to the line of sight (as seen for DLAs in the spectra of 2 QSOs close to each other as projected on the sky), as well as from kinematic signatures in the Ly$\alpha$ profile shapes (Prochaska \& Wolfe 1997) masses of $10^{10}$ -- ${\rm 10^{11}~M_{\odot}}$ are estimated for DLAs. Their cumulative comoving gas densities at ${\rm z \sim 2 - 3}$ roughly equal the local comoving densities of (stars $+$ gas) in spirals. This lead to the paradigm that DLAs are the progenitors of present-day spirals (e.g. Wolfe 1995). Recently, however, an origin of DLA absorption in starbursting dwarfs (Vladilo 1998) or LSB galaxies (Jimenez \etal 1999) is also discussed as well as an origin in subgalactic fragments bound to merge for the highest-z systems (Haehnelt \etal 1998). Attempts to optically identify DLA absorber have been remarkably unsuccessful and yielded a confusing plenitude of morphologies at low-z. Thus, the question as to the galaxy population giving rise to DLA absorption at various redshifts seems still open. 

We were surprised to see that without any adjustment or scaling our spiral models with SFHs that successfully reproduce local galaxies in terms of spectrophotometric and chemical properties and the spectral properties of high-z galaxies provide fairly good agreement with the abundance evolution of DLAs over the entire redshift range. For all elements with a reasonable number of DLA abundance data and over the entire redshift range, our Sa and Sd models bracket the data from the high and low metallicity side, respectively. We conclude that {\bf from the point of view of abundance evolution DLAs may well be the progenitors of normal spiral galaxies}. Our models bridge the gap from high-z DLA to local HII region abundances. The weak redshift evolution of DLA abundances is a natural consequence of the long SF timescales well established for disk galaxies, the scatter in the data at any redshift is fully accounted for by the range of SF timescales from early to late type spirals (see Lindner \etal 1999). While clearly our assumption of a single value for the redshift of 
galaxy formation is a poor approximation to a realistic extended period of galaxy formation, its value does not affect our conclusions as long as there are galaxies forming before z = 4. 

The comparison of model enrichment with DLA data further suggests that while at high redshift all kinds of spirals from Sa through Sd seem to give rise to DLA absorption, this changes towards low redshifts. Our Sa -- Sc models reach metallicities higher than any DLA observation as their gas-to-total-mass ratios drop below $\sim 50$ \% (including HI beyond the optical disks). We conclude that DLA galaxy samples at low redshift should be biased against early type spirals 1.) because of their low gas content  and 2.) because of their high dust content ensuing from their high metallicity. Indeed, DLA systems are observed to show systematically lower N(HI) at low-z than at high-z (Lanzetta \etal 1997). While a bias against high metallicity absorbers has been discussed by several authors as a consequence of the ensuing dust absorption that removes the QSO from magnitude limited samples (eg. Steidel \etal 1997), our models indicate an additional reason: {\bf the rise in metallicity goes with a decrease in gas content and reduces the probability for a QSO sightline to cut through a high enough column density region for DLA absorption.} 

The change with redshift that we find in the DLA galaxy population has important consequences for the spectrophotometric properties expected for the DLA absorbing galaxies. Early type spirals being intrinsically brighter locally by $\sim 2$ mag on average in B than late type spirals, our cc spectro-cosmological models predict that their absence from 
low-z DLA samples will make it as difficult to detect low-z DLA galaxies as it is for high-z DLA absorbers. In all three bands B, \R, and K, the intrinsically brighter Sa's among the DLAs at ${\rm z \sim 2 - 3}$ have about the same apparent magnitudes as the intrinsically fainter Sd's that make up the low-z DLA population. We predict ${\rm \langle B \rangle \sim 24 - 25.5,~ \langle}$ \R ${\rm \rangle \sim 24 - 24.5,~ \langle K \rangle \sim 21.4 - 22}$ both for late type spirals at z $\sim 0.5$ and for early type spirals at z $\sim 2 - 3$. 

\begin{figure}
\vspace{-6pt}
\centerline{\psfig{file=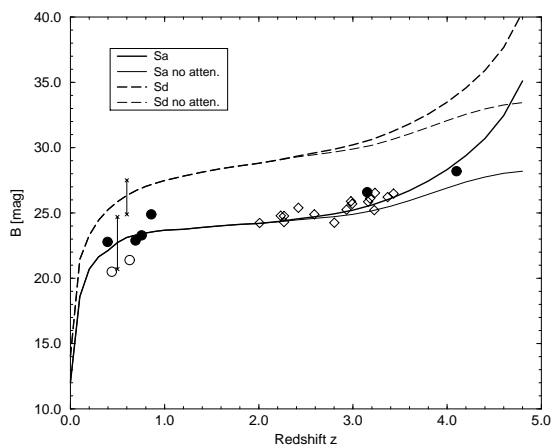,width=20pc,angle=270}}
\vspace{-5pt}
\caption{\small\baselineskip4mm Redshift evolution of B magnitude for models
 Sa and Sd w/o attenuation compared to Lyman break galaxies (Lowenthal \etal 1997) (open diamonds) and to optically identified DLA galaxies (filled circles) or galaxy candidates (open circles). The four fainter DLA galaxies at ${\rm z < 1}$ are from Steidel \etal 1994, 1995, the two high-z DLA galaxies from Djorgovski \etal 1996, 1997, the two brighter DLA galaxy {\bf candidates} are from Le Brun \etal 1997. Light vertical lines on the Sa and Sd curves indicate their brightness range in Virgo. ${\rm (H_o,\Omega_o,\Lambda_o)=(50,~1.0,~0),~z_{form}=5}$}
\vspace{-6pt}
\end{figure}

In Fig.1 we present the B-band redshift evolution of our cc spectro-cosmological models together with a sample of Lyman break galaxies and a few optically identified DLA absorbers. The Lyman break galaxies as well as the optically identified DLA 
galaxies and candidates agree with our early to intermediate type spiral models. The two DLA candidates at ${\rm z = 0.44~and~0.63}$ indeed show a bright spiral morphology (Boiss\'e \etal 1998). Note that while chemical abundance determinations do not select bright galaxies the optical identifications do. 
Very deep imaging and spectroscopy both with HST and from the ground should reveal the nature of the DLA galaxy population both at high and low redshifts. Models need to be extended to include infall, spatial resolution and dynamical evolution.   

\smallskip\noindent
{\small
{\sl Acknowledgement:} UFvA and CSM gratefully acknowledge travel support from the organisers 
without which our participation would not have been possible.} 


\vspace{-.6cm}

\begin{moriondbib}
\vspace{-.4cm}
{\small\baselineskip4mm
\bibitem{} Boiss\'e, P., Le Brun, V., Bergeron, J., Deharveng, J.-M., 1998, A\&A 333, 841
\bibitem{} Le Brun, V., Bergeron, J., Boiss\'e, P., Deharveng, J. M., 1997, A\&A 321, 733
\bibitem{} Djorgovski, S. G., 1997, in {\sl Structure and Evolution of
the IGM from QSO Absorption Lines} eds. P. Petitjean, S. Charlot, Editions Fronti\`eres, p. 303
\bibitem{} Djorgovski, S. G., Pahre, M. A., Bechtold, J., Elston, R., 1996, Nat 382, 234 
\bibitem{} Einsel, C., Fritze -- v. Alvensleben, U., Kr\"uger, H., Fricke, K. J., 1995, A\&A 196, 374
\bibitem{} Ferguson, A.~M.~N., Gallagher, J.~S., Wyse, R.~F.~G., 1998, AJ 116, 673
\bibitem{} Haehnelt, M. G., Steinmetz, M., Rauch, M., 1998, ApJ 495, 647
\bibitem{} v. d. Hoek, L. B., Groenewegen, M. A. T., 1997, A\&AS 123, 305
\bibitem{} Jimenez, R., Bowen, D. V., Matteucci, F., 1999, ApJ 514, L83
\bibitem{} Kennicutt, R.~C., 1992, ApJS 79, 255
\bibitem{} Lanzetta, K. M., Wolfe, A. M., Altan, H., \etal, 1997, AJ 114, 1337
\bibitem{} Lejeune, T., Cuisinier, F., Buser, R., 1997, A\&AS, 125, 229 $+$ 1998, A\&AS 130, 65
\bibitem{} Lindner, U., Fritze -- v. Alvensleben, U., Fricke, K. J., 1999, A\&A 341, 709
\bibitem{} Lowenthal, J.~D., Koo, D.~C., Guzman, R., \etal, 1997, ApJ 481, 673
\bibitem{} M\"oller, C.~S., Fritze -- v. Alvensleben, U., Fricke, K. J., 1997, A\&A 317, 676
\bibitem{} Oey, M.~S., Kennicutt, R.~C., 1993, ApJ 411, 137
\bibitem{} Phillipps, S., Edmunds, M. G., 1996, MN 281, 362
\bibitem{} Prochaska, J. X., Wolfe, A. M., 1997, ApJ 474, 140
\bibitem{} Vladilo, G., 1998, ApJ 493, 583
\bibitem{} Steidel, C. C., Pettini, M., Dickinson, M., Persson, S. E., 1994, AJ 108, 2046
\bibitem{} Steidel, C. C., Bowen, D. V., Blades, J. C.,
Dickinson, M., 1995, ApJ 440, L45
\bibitem{} Steidel, C. C., Dickinson, M., Meyer, D. M., \etal, 1997, ApJ 480, 568
\bibitem{} Wolfe, A.~M., 1995, in {\sl QSO Absorption Lines}, ed. G.~Meylan, Springer, p. 13
\bibitem{} Woosley, S. E., Weaver, T. A., 1995, ApJS 101, 181
\bibitem{} Zaritsky, D., Kennicutt, R.~C., Huchra, J.~P., 1994, ApJ 420, 87
}
\end{moriondbib}
\end{document}